\documentclass{PoS}

\usepackage{wrapfig}   

\title{Search for the QCD critical point at SPS energies}

\ShortTitle{Search for the QCD critical point at SPS energies}

\author{\speaker{Katarzyna Grebieszkow} for the NA49 and the NA61/SHINE
 Collaborations \\ 
\\ 
        Warsaw University of Technology\\
        E-mail: \email{kperl@if.pw.edu.pl}}

\abstract
{Lattice QCD calculations locate the QCD critical point at energies
accessible at the CERN Super Proton Synchrotron (SPS). We present average
transverse momentum and multiplicity fluctuations, as well as baryon and
anti-baryon transverse mass spectra which are expected to be sensitive
to effects of the critical point. The future CP search strategy
of the NA61/SHINE experiment at the SPS is also discussed.}

\author{The NA49 Collaboration: \\
T.~Anticic$^{23}$, B.~Baatar$^{8}$,D.~Barna$^{4}$,
J.~Bartke$^{6}$, L.~Betev$^{10}$, H.~Bia{\l}\-kowska$^{20}$,
C.~Blume$^{9}$,  B.~Boimska$^{20}$, M.~Botje$^{1}$,
J.~Bracinik$^{3}$, P.~Bun\v{c}i\'{c}$^{10}$,
V.~Cerny$^{3}$, P.~Christakoglou$^{2}$,
P.~Chung$^{19}$, O.~Chvala$^{14}$,
J.G.~Cramer$^{16}$, P.~Csat\'{o}$^{4}$, P.~Dinkelaker$^{9}$,
V.~Eckardt$^{13}$,
Z.~Fodor$^{4}$, P.~Foka$^{7}$,
V.~Friese$^{7}$, J.~G\'{a}l$^{4}$,
M.~Ga\'zdzicki$^{9,11}$, V.~Genchev$^{18}$, 
E.~G{\l}adysz$^{6}$, K.~Grebieszkow$^{22}$,
S.~Hegyi$^{4}$, C.~H\"{o}hne$^{7}$,
K.~Kadija$^{23}$, A.~Karev$^{13}$, D.~Kikola$^{22}$,
V.I.~Kolesnikov$^{8}$, E.~Kornas$^{6}$,
R.~Korus$^{11}$, M.~Kowalski$^{6}$,
M.~Kreps$^{3}$, A.~Laszlo$^{4}$,
R.~Lacey$^{19}$, M.~van~Leeuwen$^{1}$,
P.~L\'{e}vai$^{4}$, L.~Litov$^{17}$, B.~Lungwitz$^{9}$,
M.~Makariev$^{17}$, A.I.~Malakhov$^{8}$,
M.~Mateev$^{17}$, G.L.~Melkumov$^{8}$, A.~Mischke$^{1}$, 
M.~Mitrovski$^{9}$,
J.~Moln\'{a}r$^{4}$, St.~Mr\'owczy\'nski$^{11}$, V.~Nicolic$^{23}$,
G.~P\'{a}lla$^{4}$, A.D.~Panagiotou$^{2}$, D.~Panayotov$^{17}$,
A.~Petridis$^{2,\ast}$, W.~Peryt$^{22}$, M.~Pikna$^{3}$, 
J.~Pluta$^{22}$,
D.~Prindle$^{16}$,
F.~P\"{u}hlhofer$^{12}$, R.~Renfordt$^{9}$,
C.~Roland$^{5}$, G.~Roland$^{5}$,
M. Rybczy\'nski$^{11}$, A.~Rybicki$^{6}$,
A.~Sandoval$^{7}$, N.~Schmitz$^{13}$, T.~Schuster$^{9}$, 
P.~Seyboth$^{13}$,
F.~Sikl\'{e}r$^{4}$, B.~Sitar$^{3}$, E.~Skrzypczak$^{21}$, 
M.~Slodkowski$^{22}$,
G.~Stefanek$^{11}$, R.~Stock$^{9}$, C.~Strabel$^{9}$, 
H.~Str\"{o}bele$^{9}$, T.~Susa$^{23}$,
I.~Szentp\'{e}tery$^{4}$, J.~Sziklai$^{4}$, M.~Szuba$^{22}$, 
P.~Szymanski$^{10,20}$,
V.~Trubnikov$^{20}$, M.~Utvic$^{9}$, D.~Varga$^{4,10}$, 
M.~Vassiliou$^{2}$, G.I.~Veres$^{4,5}$, G.~Vesztergombi$^{4}$,
D.~Vrani\'{c}$^{7}$, 
Z.~W{\l}odarczyk$^{11}$, A.~Wojtaszek-Szwarc$^{11}$, I.K.~Yoo$^{15}$

}

\author{ \\
$^{1}$NIKHEF, Amsterdam, Netherlands. \\
$^{2}$Department of Physics, University of Athens, Athens, Greece.\\
$^{3}$Comenius University, Bratislava, Slovakia.\\
$^{4}$KFKI Research Institute for Particle and Nuclear Physics, 
Budapest, Hungary.\\
$^{5}$MIT, Cambridge, USA.\\
$^{6}$Institute of Nuclear Physics, Cracow, Poland.\\
$^{7}$Gesellschaft f\"{u}r Schwerionenforschung (GSI), Darmstadt, 
Germany.\\
$^{8}$Joint Institute for Nuclear Research, Dubna, Russia.\\
$^{9}$Fachbereich Physik der Universit\"{a}t, Frankfurt, Germany.\\
$^{10}$CERN, Geneva, Switzerland.\\
$^{11}$Institute of Physics, Jan Kochanowski University, Kielce, 
Poland.\\
$^{12}$Fachbereich Physik der Universit\"{a}t, Marburg, Germany.\\
$^{13}$Max-Planck-Institut f\"{u}r Physik, Munich, Germany.\\
$^{14}$Institute of Particle and Nuclear Physics, Charles University, 
Prague, Czech Republic.\\
$^{15}$Department of Physics, Pusan National University, Pusan, Republic 
of Korea.\\
$^{16}$Nuclear Physics Laboratory, University of Washington, Seattle, 
WA, USA.\\
$^{17}$Atomic Physics Department, Sofia University St. Kliment Ohridski, 
Sofia, Bulgaria.\\
$^{18}$Institute for Nuclear Research and Nuclear Energy, Sofia, 
Bulgaria.\\
$^{19}$Department of Chemistry, Stony Brook Univ. (SUNYSB), Stony Brook, 
USA.\\
$^{20}$Institute for Nuclear Studies, Warsaw, Poland.\\
$^{21}$Institute for Experimental Physics, University of Warsaw, Warsaw, 
Poland.\\
$^{22}$Faculty of Physics, Warsaw University of Technology, Warsaw, 
Poland.\\
$^{23}$Rudjer Boskovic Institute, Zagreb, Croatia.\\
$^{\ast}$deceased
}

\author{The NA61/SHINE Collaboration: \\
N.~Abgrall${}^{22}$,
A.~Aduszkiewicz${}^{23}$,
B.~Andrieu${}^{11}$,
T.~Anticic${}^{13}$,
N.~Antoniou${}^{18}$,
J.~Argyriades${}^{22}$,
A.~G.~Asryan${}^{15}$,
B.~Baatar${}^{9}$,
A.~Blondel${}^{22}$,
J.~Blumer${}^{5}$
% ...

}

\author{ \\
${}^{ 1}$Cape Town University, Cape Town, South Africa \\
${}^{ 2}$ETH, Zurich, Switzerland \\
${}^{ 3}$Fachhochschule Frankfurt, Frankfurt, Germany \\
${}^{ 4}$Faculty of Physics, University of Sofia, Sofia, Bulgaria \\
${}^{ 5}$Forschungszentrum Karlsruhe, Karlsruhe, Germany \\
${}^{ 6}$Institute for Nuclear Research, Moscow, Russia \\
${}^{ 7}$Institute for Particle and Nuclear Studies, KEK, Tsukuba,  
Japan \\
${}^{ 8}$Jagiellonian University, Cracow, Poland  \\
${}^{ 9}$Joint Institute for Nuclear Research, Dubna, Russia \\
${}^{10}$KFKI Research Institute for Particle and Nuclear Physics, 
Budapest, Hungary \\
${}^{11}$LPNHE, University of Paris VI and VII, Paris, France \\
${}^{12}$University of Silesia, Katowice, Poland  \\
${}^{13}$Rudjer Boskovic Institute, Zagreb, Croatia \\
${}^{14}$Soltan Institute for Nuclear Studies, Warsaw, Poland \\
${}^{15}$St. Petersburg State University, St. Petersburg, Russia \\
${}^{16}$State University of New York, Stony Brook, USA \\
${}^{17}$Jan Kochanowski University in  Kielce, Poland \\
${}^{18}$University of Athens, Athens, Greece \\
${}^{19}$University of Bergen, Bergen, Norway \\
${}^{20}$University of Bern, Bern, Switzerland \\
${}^{21}$University of Frankfurt, Frankfurt, Germany \\
${}^{22}$University of Geneva, Geneva, Switzerland \\
${}^{23}$University of Warsaw, Warsaw, Poland \\
${}^{24}$Warsaw University of Technology, Warsaw, Poland  \\

}

\FullConference{European Physical Society Europhysics Conference on High Energy Physics,
EPS-HEP 2009,\\
		 July 16 - 22 2009\\
		 Krakow, Poland}

\begin{document}

\section{Introduction and motivation}

\begin{wrapfigure}{r}{7.cm}
\vspace{-0.8cm}
\includegraphics[scale=0.38]{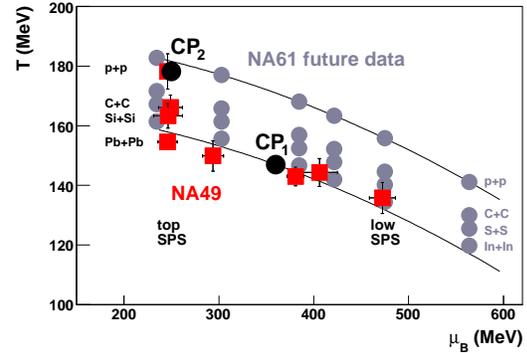}
\vspace{-0.9cm}
\caption[]{Chemical freeze-out points in NA49 (red) and those
expected in NA61 (violet). $CP_{1}$ and $CP_{2}$ were considered in
NA49 as possible locations of the critical point: $\mu_B (CP_1)$
from lattice QCD calculations \cite{fodor_latt_2004} and $CP_2$
assuming that the chemical freeze-out point of p+p data at 158$A$ GeV
may be located on the phase transition line.}
\label{cp1cp2}
\vspace{-0.5cm}
\end{wrapfigure}

Theoretical calculations suggest that the critical point (CP) of 
strongly interacting matter may be accessible in the SPS energy range 
\cite{fodor_latt_2004}. We studied event-by-event average 
$p_T$ and multiplicity fluctuations, as well as transverse mass
spectra of baryons and anti-baryons which are suggested observables
sensitive to effects of the CP in ultra-relativistic heavy ion collisions. 

The effects are expected to be maximal when freeze-out happens
near the critical point. The position of chemical freeze-out point in 
the $(T - \mu_B)$ diagram can be varied by changing the energy and the 
size of the colliding system (Fig. \ref{cp1cp2}). Therefore we analyzed
in NA49 \cite{na49_nim} the energy dependence of the proposed CP sensitive 
observables for central Pb+Pb collisions (beam energies 20$A$-158$A$ 
GeV), and their system size dependence (p+p, C+C, Si+Si, and Pb+Pb) at 
the highest SPS energy.

\section{Event-by-event average $p_T$ and multiplicity fluctuations}

Enlarged event-by-event fluctuations of multiplicity N and mean $p_T$ were 
suggested as a signature of the critical point \cite{SRS}. The NA49 
experiment used the $\Phi_{p_T}$ correlation measure \cite{ 
fluct_size, fluct_energy} and the scaled variance of the multiplicity 
distribution $\omega$ \cite{omega_size, omega_energy} to study average 
$p_T$ and $N$ fluctuations, respectively. For $\omega$, we selected very 
central collisions only (1\% most central) due to its strong 
dependence on fluctuations of the number of participants $N_{part}$. 

The energy ($\mu_B$) dependence of $\Phi_{p_T}$ and $\omega$ together 
with predictions for $CP_1$ were presented at this conference (see also 
\cite{kg_qm09}). The NA49 data show no significant peak in the energy 
dependence of $\Phi_{p_T}$ and $\omega$ at SPS energies thus providing no 
indications of the critical point at $CP_1$ (see Fig.~\ref{cp1cp2}).

\begin{figure}
\centering
\vspace{-0.3cm}
\includegraphics[width=.8\textwidth]{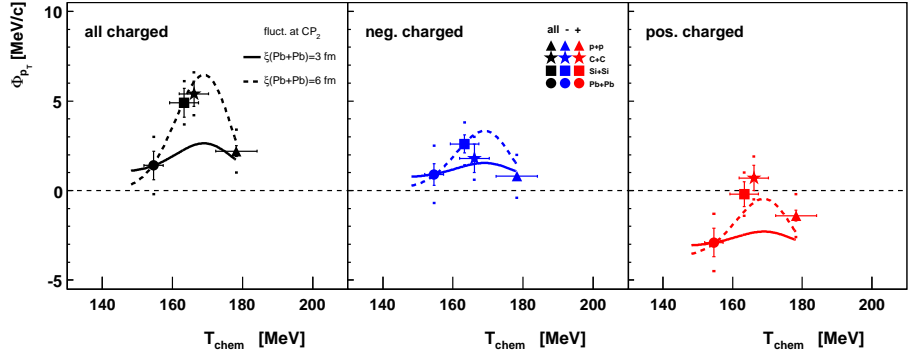}
\vspace{-0.5cm}
\caption[]{System size dependence of $\Phi_{p_T}$ at 158$A$ GeV 
(forward rapidity, NA49 azimuthal angle acceptance) showing 
results from p+p, semi-central C+C (15.3\%) and Si+Si (12.2\%), and 5\% 
most central Pb+Pb collisions \cite{fluct_size}. Lines correspond to
$CP_2$ predictions (see text) shifted to reproduce the $\Phi_{p_T}$ 
value for central Pb+Pb collisions.}
\label{fiptT}
\end{figure}

\begin{figure}
\centering
\vspace{-0.3cm}
\includegraphics[width=.8\textwidth]{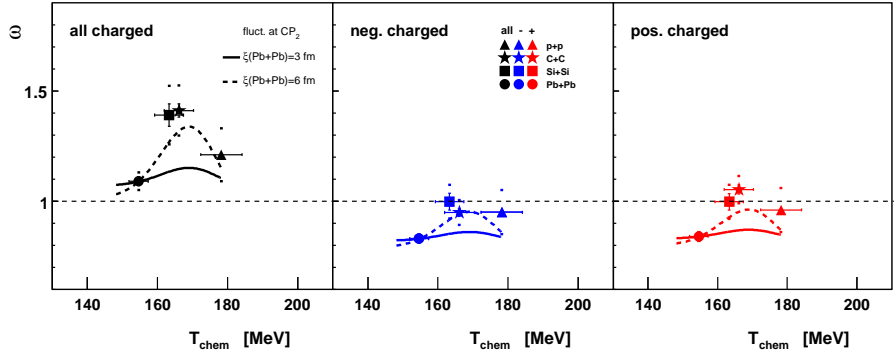}
\vspace{-0.5cm}
\caption[]{System size dependence of $\omega$ at 158$A$ GeV (forward 
rapidity, NA49 azimuthal angle acceptance) for the 1\% most 
central p+p \cite{omega_size}, C+C and Si+Si \cite{benjaminPhD}, and 
Pb+Pb collisions \cite{omega_energy}. Lines
correspond to $CP_2$ predictions (see text) shifted to reproduce the 
$\omega$ value for central Pb+Pb collisions. }
\label{omegaT}
\end{figure}

Figures \ref{fiptT} and \ref{omegaT} present the system size
($T_{chem}$\footnote{$T_{chem}$ values were taken from fits of the 
hadron gas model \cite{beccatini} to particle yields.}) dependence of 
$\Phi_{p_T}$ and $\omega$. The lines correspond to predictions for
$CP_2$ (see Fig. \ref{cp1cp2}) with estimated magnitude of the effects
\footnote{Predicted magnitudes include corrections by NA49 due
to the limited rapidity range (forward-rapidity) and azimuthal
angle acceptance of the detector.}
for $\Phi_{p_T}$ and $\omega$ at $CP_2$ taken from Ref.\cite{SRS, 
MS} assuming correlation lengths $\xi$ decreasing monotonically
with decreasing system size: a) $\xi$(Pb+Pb) = 6 fm and  $\xi$(p+p) = 2 
fm (dashed lines) or b) $\xi$(Pb+Pb) = 3 fm and  $\xi$(p+p) = 1 fm 
(solid lines). The width of the enhancement due to CP in the ($T, 
\mu_B$) plane is based on Ref. \cite{hatta} and taken as $\sigma (T) 
\approx 10$ MeV. 
A maximum of $\Phi_{p_T}$ and $\omega$ is observed for C+C 
and Si+Si interactions at the top SPS energy. It is two times higher for all 
charged than for negatively charged particles, as expected for the 
effect of the CP \cite{SRS}. Results presented in Figs. \ref{fiptT} and 
\ref{omegaT} suggest that the NA49 data are consistent with $CP_2$ 
predictions.

\begin{figure}    
\centering
\vspace{-0.3cm}
\includegraphics[width=.8\textwidth]{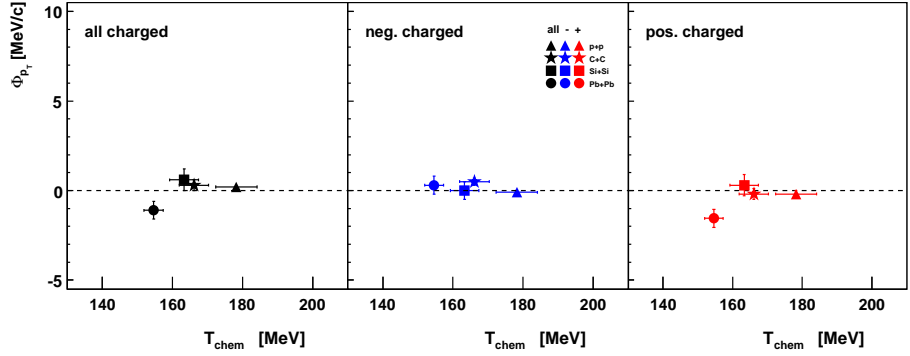}
\vspace{-0.5cm}
\caption[]{The same as Fig. \ref{fiptT} but high $p_T$ region shown 
($0.5 < p_T < 1.5$ GeV/c).}
\label{high_pt}
\end{figure}

\begin{figure}    
\centering
\vspace{-0.3cm}
\includegraphics[width=.8\textwidth]{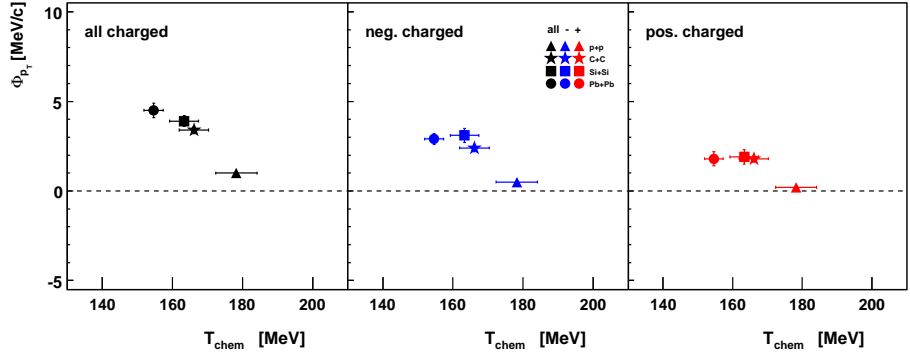}
\vspace{-0.5cm}
\caption[]{The same as Fig. \ref{fiptT} but low $p_T$ region shown 
($0.005 < p_T < 0.5$ GeV/c).}
\label{low_pt}
\end{figure}

It is expected that fluctuations due to the CP originate mainly 
from low $p_T$ pions \cite{SRS}. Therefore the NA49 analysis of $\Phi_{p_T}$ 
was performed also for two separate $p_T$ regions (Figs. \ref{high_pt} 
and \ref{low_pt}). Indeed, the high $p_T$ region shows fluctuations 
consistent with zero (Fig. \ref{high_pt}) and correlations are observed 
predominantly at low $p_T$ (Fig. \ref{low_pt}). However, in low $p_T$ 
region, data do not show a maximum of $\Phi_{p_T}$, but a continuous
rise towards Pb+Pb collisions. The origin of this behavior is currently 
being analyzed (short range correlations are considered).

\section{Transverse mass spectra of baryons and anti-baryons}

It was suggested \cite{Askawa} that the critical point serves as an 
attractor of hydrodynamical trajectories in the $(T, \mu_B)$ phase
diagram. This was conjectured to lead to a decrease of the anti-baryon 
to baryon (${\bar{B}} / {B}$) ratio with increasing transverse momentum. 
The ${ \bar{p}} / {p}$, ${ \bar{\Lambda}} / {\Lambda}$, and ${ 
{\bar{\Xi}}^{+}} / {\Xi^{-}}$ ratios versus reduced transverse mass 
$m_T-m_0$ were studied by the NA49 experiment \cite{kg_qm09} and 
presented at this conference. The slopes of all three ${\bar{B}} / 
{B}$ ratios show no significant energy dependence, thus implying 
that transverse mass spectra of $B$ and ${\bar{B}}$ do not provide evidence 
for the critical point in the SPS energy range.

\section{Summary of NA49 results and strategy of NA61/SHINE}

The energy dependence of average $p_T$ and multiplicity fluctuations, 
and ratios of the anti-baryon/baryon transverse mass spectra in central
Pb+Pb collisions provide no indications of the critical point. The system 
size dependence at 158$A$ GeV exhibits a maximum of mean $p_T$ and 
multiplicity fluctuations in the complete $p_T$ range (consistent with 
$CP_2$ predictions) and an increase (from p+p up to Pb+Pb) of mean $p_T$ 
fluctuations in the low $p_T$ region. The low $p_T$ region will be 
carefully analyzed for the effects of short range correlations on 
$\Phi_{p_T}$ and $\omega$.

A detailed energy and system-size scan is necessary to establish the 
existence of the critical point. Therefore
the CP search will be continued by the NA61/SHINE \cite{shine} 
experiment which is based on the upgraded NA49 detector. We plan to perform 
a two-dimensional scan with lighter ions (p+p, C+C, S+S, In+In) in a broad 
beam energy range (10$A$ - 158$A$ GeV). The hypothetical chemical 
freeze-out points in the NA61 experiment are presented in Fig. 
\ref{cp1cp2}.Together with existing NA49 data the scan may help to locate the 
QCD critical point in the $(T, \mu_B)$ phase diagram.

\end{document}